\documentclass{aa}

\usepackage{graphicx}
\usepackage{txfonts}
\usepackage{soul}
\usepackage{color}

\usepackage[pdfpagelabels=false]{hyperref}	
\hypersetup{colorlinks=true,linkcolor=blue,citecolor=blue,filecolor=blue,urlcolor=blue,}
\usepackage{multicol}
\usepackage{subcaption}

\begin{document}

   \title{The progenitor galaxies of the intracluster light}

   \author{Rebecca J. Mayes, 
          \inst{1}
        Facundo A. G\'omez
        \inst{1}
          \and
          Antonela Monachesi\inst{1}
          }

   \institute{Departamento de Astronomía, Universidad de La Serena, Avda. Raúl Bitrán Nº 1305, La Serena, Chile\\
              \email{rebecca.jane@userena.cl}
             \thanks{}
             }

   \date{Received x; accepted y}

\abstract{Intracluster Light (ICL) is a diffuse stellar component of galaxy groups and clusters, bound to the structure's potential rather than to individual galaxies and primarily formed through the stripping and mergers of intermediate-mass galaxies. Its formation and evolution is strongly linked to the central galaxies of groups and clusters, typically the Brightest Cluster Galaxies (BCGs) in clusters.}{To determine the properties of the progenitor galaxies of the ICL of clusters and groups at redshift z = 0 and compare them with the progenitors of their BCGs/BGGs (Brightest Cluster Galaxies/Brightest Group Galaxies).}{We select 127 clusters and groups (structures) in the hydrodynamic Illustris-TNG100 simulation above a mass of 10\textsuperscript{13}~\(\textup{M}_\odot\). We divide the ICL from the BCG by applying a surface brightness cut at the Holmberg radius of 26.5 mag/arcsec\textsuperscript{-2}, where star particles within this radius are defined as being attached to the BCG, and outside, the ICL. We then determine the masses and mass contributions of progenitor galaxies of the ICL and BCG, for comparison.}{We find that the number of progenitors of the ICL increases with structure mass, and the ICL has more progenitors than the BCG. The ICL has a larger percentage of material contributed by low-mass galaxies. A large fraction of the ICL was contributed by quenched galaxies, with this fraction increasing with galaxy mass. The mean mass radial gradient of ex-situ material in structures for the ICL and the BCG+ICL is generally negative and increases with structure mass, indicating that material from more massive progenitors is found towards the centres of clusters. The mean mass radial gradient is significantly impacted by structure formation history.}{ICL progenitor properties are impacted by structure history and are therefore highly individual to each system. At a given structure mass, there is diversity in the properties of the progenitors that formed the ICL. Although the BCG and the ICL co-evolve, the BCG has fewer, more massive progenitors.} 

    \titlerunning{Coevolution of ICL and BCGs}
    \authorrunning{Rebecca J. Mayes, Facundo A. G\'omez
        \and Antonela Monachesi\inst{1}}
   \maketitle
%

\section{Introduction}
The intracluster light (ICL) in galaxy clusters, also called intragroup light (IGL) in groups, is a faint, diffuse stellar component bound to the cluster potential rather than to individual galaxies. It was first theorised and later discovered in the Coma cluster \citep{Zwicky1937, Zwicky1951}. The ICL extends for hundreds of parsecs beyond the structure centre and makes up a significant fraction of the stellar component of galaxy clusters, between 5-50 per cent, which can vary depending on how the ICL is defined \citep[e.g.][]{Mihos_2015ICL, contini2021b, montes2022}. Many theoretical and observational studies have examined the ICL and its link to cluster formation and evolution (see \citealt{contini2024} for a recent review).

Because the ICL is comprised of stars influenced only by the potential well of the dark matter distribution, it has been found to be a possible luminous tracer of the underlying dark matter distribution, in observations \citep[e.g.][]{Montes2019, SampaioSantos2021, Diego_2023} and simulations \citep[e.g.][]{Alonso_Asensio_2020, Yoo_2022, Yoo_2024}. 
However, it may be biased, particularly in unrelaxed systems, because it differs in specific-energy distribution, orbital anisotropy, and radial concentration from the dark matter \citep{Butler_2025}. The ICL grows in tandem with the brightest cluster galaxy (BCG), the galaxy located closest to the minima of the potential well in host clusters \citep{Jones1984}. Due to their locations, BCGs are strongly linked to the evolution of their host clusters, and they exhibit properties distinct from those of typical elliptical galaxies. BCGs usually have higher velocity dispersions and larger radii than non-BCG ellipticals \citep{Jones1984, VonDerLinden2007}, more diverse colour profiles \citep{Bildfell_2008} and are more likely to host radio-loud AGN \citep{chu2022}, among other distinct properties.

The ICL was first identified as an extended component of the BCG halos \citep[e.g.][]{Matthews1964}. Despite this, the BCG and the ICL have been shown to present extremely distinct properties. The ICL is older, bluer, and metal-poorer than the BCG. \citep{Morishita2017, Montes2018, DeMaio2018, Contini2019, mayes2025}, and the accreted portions of both  show distinct differences. The BCG grows in a two-phase process through early in-situ growth and major mergers, and then a second, later phase of major and minor mergers \citep[e.g.][]{DeLucia2007, Naab2009, Laporte2012}, while the ICL is predominantly produced at lower redshifts through tidal stripping of intermediate mass satellites \citep{Monaco2006, Murante2007, Somerville2008,Guo2011,Martel2012,contini2014,Contini_2018, Brown2024, mayes2025}. The in-situ component of the ICL is much smaller than that of the BCG \citep{mayes2025}, and ongoing star formation has not been found in observations \citep{Melnick_2012}. The only observational evidence of in-situ star formation in the ICL is from \citet{Barfety_2022}, who observe in-situ star formation in the ICL in a cluster at z=1.7. Observations of jellyfish galaxies demonstrate that stars can form in situ within ram-pressure-stripped gaseous tails embedded in the intracluster medium. While this channel is not typically quantified separately in ICL studies, it provides a possible mechanism for the direct formation of intracluster stars and may contribute to the in-situ component of the ICL \citep{Gullieuszik_2020, George_2022, Moretti2026}.

The progenitors that contribute to the ICL are typically satellites that are undergoing tidal disruption \citep{Rudick2009, Rudick_2011, Montes_2014, Contini_2018,DeMaio2018,Montes2018,contini2024}, while the progenitors that contribute to the BCG are more likely to have been fully disrupted \citep{mayes2025}. Simulations have found that the main contributors to ICL formation should be high-mass (>10\textsuperscript{10}~\(\textup{M}_\odot\)) galaxies \citep{Contini2019,contini2024,Chun2023} and typically a small number of progenitors contribute the majority of the material in the ICL \citep{Cooper20150}. For example, \citet{Harris_2017} simulated a Fornax-like cluster and found >60 per cent of the z = 0 ICL formed from just two massive objects. \citet{Ahvazi2024} studied ICL formation in the TNG50 simulation and found that half the ICL across considered systems originated in galaxies with stellar masses between 10\textsuperscript{10} and 10\textsuperscript{11}~\(\textup{M}_\odot\). The ICL may also form via pre-processing, in which the ICL forms as IGL in the group environment and later merges with the host cluster. Up to 30 per cent of the ICL is formed by this mechanism \citep{contini2014}. However, in this scenario, the ICL formed earlier through mergers or stripping and was subsequently accreted, making it a sub-channel rather than a primary mechanism. Ongoing questions remain about the origins of galaxies stripped to form the ICL, as well as how their properties may vary depending on the environment and the different merger histories a cluster may follow.

It is difficult to define the ICL precisely and to separate it from the BCG, both in observations and in simulations. Several mechanisms can be used to divide the ICL and the BCG, often yielding different results \citep[e.g.][]{Brough2023}. In observations, the ICL is typically first separated from background galaxies using an isophotal limit to cut off light from satellite galaxies. Subsequently, in the inner regions, the ICL is separated from the BCG using profile decomposition \citep[e.g.][]{Zibetti2005} or two-dimensional profile fitting to model the surface brightness profiles of BCGs \citep[e.g.][]{Gonzalez2005}. In simulations, different mechanisms for separating the ICL and the BCG in the inner regions include: (1) A simple radius cut, usually of 30 kpc or 100 kpc, where all stellar material outside the cut, and not bound to any satellite galaxies, is defined as ICL. (2) A cut based on the surface brightness profile of the BCG, for example, at the Holmberg radius of 26.5 mag/arcsec\textsuperscript{-2} \citep{Rudick_2011, mayes2025}. (3) 3-component surface brightness profile fitting that models the BCG, the BCG+ICL region and the ICL, and also allows for a characterisation of the transition zone between the two \citep[e.g.][]{Janowiecki_2010,iodice2017, kravtsov2018, Zhang2019}. Triple Sersic profile fitting is generally used for this method \citep{Spavone_2018, spavone2024}. An additional challenge is that, while theoretical models can more precisely separate the ICL, observational techniques must still be used to compare simulation and observational results fairly. Due to the difficulty of separating the BCG and the ICL, many studies have instead adopted the more straightforward approach of treating the two as a single system \citep{kravtsov2018,DeMaio2018,Zhang2019,demaio2020}. Note that some studies (for example, those using triple sersic fitting) also choose to separate the transition regime between the ICL and BCG (also known as the stellar halo of the BCG). In contrast, others simply divide the BCG and the ICL and consider the transition region to belong to either. Because of factors such as these, ICL properties can vary across studies.

A key question is how the ICL and BCG co-evolve or evolve differently, depending on the properties of the progenitor galaxies that are accreted to become the ICL and BCG. In \citet{mayes2025}, hereafter referred to as Paper 1, we compared the properties and progenitors of the ICL and the BCG in the Illustris-TNG100 simulation and found that the ICL and the BCG coevolve but have distinct formation histories and properties, with the ICL being primarily composed of material from stripped or merged intermediate-mass galaxies, while the BCG is composed of more massive merged galaxies and has a higher in situ fraction. In this paper, we expand on our previous work and aim to determine the numbers, masses, and contributions to the ICL of the progenitor galaxies that make up the ICL and the BCG, and how these vary across host structures and their properties. This paper is organised as follows. Section 2 describes the simulations used and the ICL analysis methods. Section 3 analyses the number of significant progenitors that contribute to the ICL and BCG, and Section 4 analyses the top 5 progenitors. Section 5 investigates the fraction that high-mass progenitors make up of the ICL and BCG. Section 6 discusses the formation of the ICL by preprocessing. Section 7 investigates the mass gradients of the BCG+ICL over structure masses. Section 8 summarises the main conclusions.

Note that, for simplicity, although we study diffuse light in both groups and clusters throughout this paper, we will refer to the diffuse light as ICL and the central galaxy as the BCG, even in the context of groups, rather than as IGL and the BGG. When referring to groups and clusters collectively, we will use the term "structures".


\section{Methods}
\label{section:methods}
In this section, we give an overview of the IllustrisTNG-100 simulations used to analyse the ICL and BCG and the methods used for selection and analysis. Note that many of the methods were previously covered in Paper 1, and we recommend referring back to that paper for more in-depth coverage.

\subsection{Overview of simulations}
We use data from The Next Generation Illustris Project (IllustrisTNG), \citep{Marinacci2018,Naiman2018,Nelson2018,Nelson2019,Pillepich2018b,Pillepich2019,Springel2018}. The IllustrisTNG simulations are an improved version of the original Illustris simulations \citep{Vogelsberger2013,Genel_2014,Torrey_2015,Nelson_2015} with more realistic galaxy stellar masses and sizes, and galaxy groups and clusters with gas fractions that are in better agreement with observations, and significantly improve on colour-bimodality \citep[e.g.][]{Pop2022}. TNG uses the moving-mesh code \textsc{AREPO} \citep{Springel_2010} to follow the gas dynamics coupled with dark matter using a quasi-Lagrangian treatment within the $\Lambda$CDM paradigm. Dark matter halos in the simulation are identified using a Friends-of-Friends algorithm (FoF, \citet{Davis1985}), which selects groups of particles separated by a linking length equal to or smaller than 0.2 times the mean inter-particle separation. FoF groups are generally referred to as halos. Secondly, the SUBFIND algorithm \citep{Springel2001,Dolag2009} is used to identify subhalos as gravitationally bound substructures within FoF halos. Throughout this paper, M\textsubscript{200} denotes the mass measured within the halo radius R\textsubscript{200} that encloses an average over-density of 200 times the critical density of the universe, while {\mbox{$M^{*}_{200}$} refers to the stellar mass within the same radius. 

The simulation includes boxes of 50, 100, and 300 Mpc per side, corresponding to TNG50, TNG100, and TNG300. Each of them consists of 100 snapshots from z = 20 to z = 0. Twenty snapshots are full, while the remaining eighty are mini. Both encompass the entire volume, but mini snapshots contain only a subset of the data available. Each box runs three simulations at different resolution levels. TNG includes new physical processes, such as magnetohydrodynamics, as well as prescriptions for black hole formation, growth, and multimode feedback, stellar evolution, and chemical enrichment \citep{Pillepich2018b}. TNG uses cosmology parameters derived from The \citet{Planck2016} in which $\Omega$\textsubscript{m} = 0.31, $\Omega$\textsubscript{$\Lambda$}=0.69 $\Omega$\textsubscript{b} = 0.0486 and h = 0.677.

In this work, we use TNG100 because it includes a population of high-mass galaxy clusters balanced with a higher resolution than TNG300. TNG100 is the middle-resolution baryonic run of the IllustrisTNG suite and has a volume of approximately 110\textsuperscript{3} Mpc\textsuperscript{3} and 2 × 1820\textsuperscript{3} resolution elements, with an average mass per baryonic
particle of 1.4~$\times$~10\textsuperscript{6}~\(\textup{M}_\odot\), and dark matter mass of  7.5~$\times$~10\textsuperscript{6}~\(\textup{M}_\odot\) per particle. Illustris-TNG100 hosts 127 clusters and groups above $M_{200}$ \mbox{10\textsuperscript{13}~\(\textup{M}_\odot\)}, ranging from 10\textsuperscript{13}-2.6~$\times$~10\textsuperscript{14}~\(\textup{M}_\odot\) including 10 clusters with mass \mbox{>10\textsuperscript{14}~\(\textup{M}_\odot\)}. We define a structure's state based on the 3D offset between the BCG´s most bound particle and the gas centre of mass. Structures are disturbed when the offset is larger than $0.4 \times \text{R}_{200}$, relaxed when the offset is $<0.1$ R$_{200}$, and in an intermediate state when the offset is between $0.4 \times \text{R}_{200}$ and $0.1$ R$_{200}$ \citep{Aldas_2023}. Structure concentration in TNG100 is defined as $c=R\textsubscript{200c}/R\textsubscript{s}$, with $R\textsubscript{s}$ calculated by fitting an 
NFW profile to the dark matter density profile \citep{Anbajagane_2021}.

\subsection{Selection of ICL}
Our selection of ICL is the same as that used in Paper 1. Here we briefly summarise the methods used.

To begin, we select all particles bound to central subhalos. We then rotate central subhalos using the inertia tensor to ensure that we select galaxies with the same face-on orientation. We divide the BCG and ICL with a surface brightness cut at the Holmberg radius \citep{holmberg1958} of 26.5 mag/arcsec\textsuperscript{-2} using a 1D r-band surface brightness plot, along the semi-major axis, where star particles within this radius are defined as being attached to the BCG and particles outside this radius are part of the ICL. Note that previous works using this cut have been defined in the V-band \citep{Rudick_2011}, while we use the $r-$band. The use of the $r-$band (or redder filters) is common in the analysis of stellar halos, since this component is dominated by old stellar populations that emit more strongly at longer wavelengths. We have verified that adopting a V-band surface brightness cut does not significantly affect our results.

\subsection{Origins of star particles}

We make use of the stellar assembly catalogues described in \citet{Rodriguez_Gomez2016} to divide ex-situ and in-situ particles. After selecting the ex-situ star particles, we next determine the masses of the original galaxies these particles were a part of. This is done through the following steps:
\begin{enumerate}
    \item Track back past snapshots for all particles bound to the central subhalo at z = 0 and determine the snapshot in which each star particle is no longer attached to the central subhalo.
    \item Determine what subhalo each particle is within for the snapshots selected in step (1), where the particle is no longer attached to the BCG.
    \item Replace anomalous subhalos (subhalos with no merger history that are fragments or clumps in already-formed galaxies) with those in the latest snapshot where they are located outside R\textsubscript{200}.
    \item Track the subhalos found at steps (2) and (3) back through earlier snapshots and determine the maximum stellar mass of all origin subhalos.
\end{enumerate}

Once this process is complete, we will have a final sample of subhalos that contributed to the growth of the ex-situ component of the ICL and BCG, along with their masses. Additionally, we have a map that connects each z=0 stellar particle ID to its subhalo ID at peak mass. Note that we consider diffuse light in both the group and cluster environments, but for simplicity, we refer to the diffuse light component as only ICL rather than IGL (intragroup light). Throughout this paper, we will refer to clusters and groups as structures.

\subsection{Preprocessing}
Preprocessing occurs when galaxies undergo environmental transformation, such as quenching, before they are accreted into a main cluster. In the context of ICL, preprocessing can have slightly different definitions. It can refer to (1) Stars that became unbound from their host galaxies during cluster infall. (2) ICL that was already diffuse within a group being accreted into a cluster or (3) ICL originating in galaxies quenched before they lost their stars to form ICL \citep[e.g.][]{contini2021b}. Different papers use different definitions, which can yield different estimated fractions. We define preprocessed ICL as that formed from progenitor galaxies that were quenched prior to cluster infall and stripping/merger, similar to point (3) above, but note that this may lead to differences when making comparisons with other studies \citep{Pallero2019}.

To determine whether an accreted subhalo has been quenched, we first find it at the point it crosses $R_{200}$. At that point, we calculate the specific star formation rate, sSFR = SubhaloSFR/galaxy stellar mass. A subhalo is considered preprocessed if it meets the criterion that $\rm sSFR(z) < 1/(6 \times t_H(z))$  \citep{Tacchella_2019}, where $t_H(z)$ is the Hubble time $1/H$ at a given redshift. This corresponds to approximately $log(\rm sSFR \times yr) = -11$ at redshift, z = 0. Note that, due to resolution limitations in IllustrisTNG, a number of subhalos in the simulation are artificially assigned SFR = 0 in the simulation output. While the specific sSFR of these subhalos is unknown, they are expected to be quenched regardless, and so are included in our quenched fraction \citep{Donnari2019, Donnari2021}.

\section{Number of significant progenitors}
\label{section:NICLprogs}

One major question regarding the formation of the ICL concerns the mass contributions of individual progenitors. In this section, we investigate the significant progenitors that make up the ICL and BCG. Progenitor galaxies can be completely disrupted or still survive as satellites. The mass of progenitor galaxies is defined as the mass within twice the stellar half-mass radius, taken from the snapshot of the progenitor's maximum mass before stripping. Note that this may underestimate galaxy masses relative to the subfind mass; however, it is preferred because outermost bound stars found by subfind may overlap with the ICL \citep{Pillepich2019}. We do not impose a minimum mass threshold on our progenitor galaxies, but note that galaxies with mass M\textsubscript{*}~<~1~$\times$~10\textsuperscript{7}~\(\textup{M}_\odot\) only make up approximately 0.1 per cent of the ICL and BCG.

Fig.~\ref{fig:sigprogmassICL} plots the number of significant progenitors that make up more than 90 per cent of the ICL against the structure total mass, henceforth referred to as $N_{90, ICL}$. $N_{90, ICL}$ ranges from 5 to 117, with a mean of 28.3. The linear fit of $N_{90, ICL}$ to structure mass has a slope of $46.1 \pm 3.1$, showing that the number of significant progenitors increases strongly with structure mass. \citet{Ahvazi2024} investigated the formation of the ICL in TNG50 and found that the number of progenitors contributing to $N_{90}$ ranged from a few to a few tens, growing with structure mass, consistent with our results.

To investigate how structure and ICL properties vary depending on $N_{90, ICL}$, we first select structures below a mass of 10\textsuperscript{13.5}~\(\textup{M}_\odot\), as in this mass range there is minimal variation in the number of significant progenitors with structure mass. In this mass range, the mean $N_{90, ICL}$ is $19.5 \pm 9.0$. Therefore, we select structures with $N_{90, ICL} >$ 30, which consists of 11 structures and structures with $N_{90, ICL} <$ 10, which consists of 12 structures for comparison. Note that this selection requires caution when comparing these results to those from clusters, as ICL formation pathways can differ between clusters and groups \citep{mayes2025}. 
\begin{figure}

	\includegraphics[width=\columnwidth]{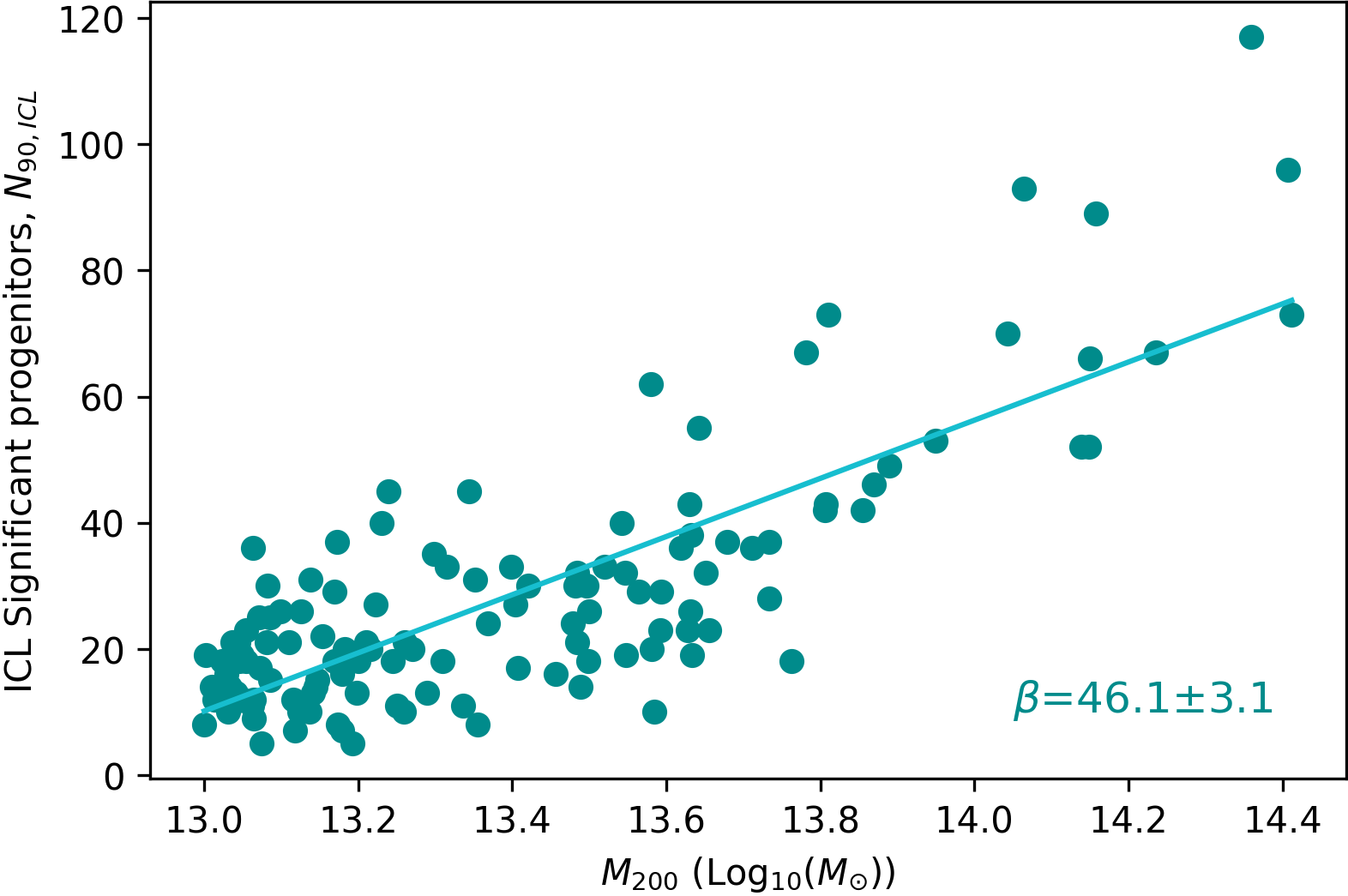}
    \caption{The number of significant progenitors that make up 90 per cent of the ICL against structure total mass}
    \label{fig:sigprogmassICL}
\end{figure}
We find that low-mass structures with $N_{90, ICL} > $30 are equally likely to be either relaxed or disturbed at 27 per cent (3/11) each, and slightly more likely to be in an intermediate state, at 45 per cent (5/11). Low mass structures with $N_{90, ICL} < $10 were slightly more likely to be disturbed, at 42 per cent (5/12), somewhat less likely to be relaxed, at 25 per cent (3/12), and moderately likely to be in an intermediate state, at 33 per cent (4/12). The mean lookback time since the last major merger for the structures with $N_{90, ICL} < $10, is $4.0 \pm 3.0$ Gyr, while structures with $N_{90, ICL} > $30 had a mean lookback time since the last major merger of $7.1 \pm 2.7$ Gyr. This likely indicates that ICL with only a few significant progenitors is dominated by material from major mergers or the merger of a recent massive galaxy, whereas those with many significant progenitors have not experienced a recent accretion of a single massive galaxy but instead ongoing accretion of multiple intermediate-mass galaxies. 

Low-mass structures with $N_{90, ICL} < $10 were more likely to share the most significant progenitor with the BCG. 58 per cent (7/12) shared the most significant progenitor, compared with only 27 per cent (3/11) of those with $N_{90, ICL} > $30. The most significant progenitor of the BCG is usually a major merger, indicating that these structures are those in which the ICL has been affected by such an event. Low mass structures with $N_{90, ICL} > $30 were more likely to be dominated by low-mass progenitors. Progenitors with mass M\textsubscript{*}~<~1~$\times$~10\textsuperscript{10}~\(\textup{M}_\odot\) made up $49 \pm 17$ per cent of the material in the ICL, in contrast to the structures with $N_{90, ICL} < $10, where the lower mass progenitors made up only $25 \pm 8$ per cent of the material of the ICL. The mean progenitor mass of the low mass structures with $N_{90, ICL} > $30 was $log(M_*) =$ $10.26 \pm 0.27$ \(\textup{M}_\odot\), while the low mass structures with $N_{90, ICL} < $10 had a mean progenitor mass of $log(M_*) =$ $10.64 \pm 0.22$ \(\textup{M}_\odot\). This would be expected as structures with $N_{90, ICL} > $30 are unlikely to have accreted many massive galaxies, while structures with $N_{90, ICL} < $10 are dominated by a few massive galaxies. 

As expected, the fraction that the most significant progenitor made up of the ICL was higher for low-mass structures with $N_{90, ICL} < $10, making up $44 \pm 16$ per cent of the ICL, while low-mass structures with $N_{90, ICL} > $30 had a most significant progenitor contribution of $19 \pm 8$. The gap was even larger for the fraction of the ICL composed of the two most significant progenitors, ranging from $60 \pm 11$ per cent to $31 \pm 10$ per cent. \citet{Cooper2010} investigated the halos of Milky Way mass galaxies by applying the GALFORM semi-analytic model to the Aquarius project. They found a typical $N_{90}$ of 1-5, with some galaxies having up to 10. A single satellite often contributed as much as 50 per cent of the halo mass. \citet{Monachesi2019} used the Auriga simulation suite to study the stellar halos of Milky Way mass galaxies and found that $N_{90}$ varied from 1 to 14, with a median of about 6.5. Galaxies dominated by a few massive progenitors tended to have more massive accreted halos. This seemingly would contradict our results (as we find that more massive structures have a larger number significant progenitors); however, there are differences between how Milky Way mass and more massive central galaxies evolve \citep{Behroozi2019}, which is discussed in more depth in Section \ref{section:top5frac}.

The number of progenitors affected the pathways by which the ICL forms. For low-mass structures with $N_{90, ICL} < $10, the percentage of the ICL that comes from progenitor galaxies that have completed their mergers was more significant, making up $63 \pm 16$ per cent, while those with $N_{90, ICL} > $30 had a completed mergers fraction of $42 \pm 11$ per cent. Correspondingly, low-mass structures with $N_{90, ICL} < $10 had a lower contribution of ICL originating in the stripping of still surviving galaxies, albeit within uncertainties of $25 \pm 18$ to $40 \pm 10$ for those with $N_{90, ICL} > $30. The in-situ component was not affected.

These results suggest that the number of significant progenitors the ICL hosts is closely tied to its formation history. Low-mass structures with less than 10 significant progenitors dominating the ICL are those that have more recently undergone completed mergers of massive galaxies, making the structure more likely to be in a disturbed state. In these cases, the ICL and the BCG are also more likely to share their most significant progenitors. While major mergers to the BCG are not generally the primary contributor to the ICL \citep{demaio2015, mayes2025}, selecting structures in which the ICL is dominated by fewer than 10 significant progenitors may disproportionately select structures in which a recent major merger is affecting ICL growth. In contrast, structures in which the ICL is composed by more than 30 significant progenitors are more likely to be those in which the dominant contributor to the ICL is the ongoing stripping of intermediate-mass galaxies. In these structures the ICL was less likely to share progenitors with the BCG, the structures had a longer lookback time since a major merger, and had a lower mean progenitor mass than structures dominated by a few progenitors \citep{mayes2025}. 

We also explore these relations for structures above a mass of 10\textsuperscript{13.5}~\(\textup{M}_\odot\). We note that this mass range has fewer structures over a larger mass range, and thus our results are statistically less significant when compared with the low mass structure sample. In this mass range, the mean $N_{90, ICL}$ is $45.7 \pm 23.6$. Therefore, we select 7 structures with $N_{90, ICL} >$ 69 and 8 structures with $N_{90, ICL} <$ 23 for comparison. High mass structures with $N_{90, ICL} > $69 are equally likely to be in a disturbed or intermediate state at 43 per cent (3/7) each, and slightly less likely to be in a relaxed state at 14 per cent (1/7). High mass structures with $N_{90, ICL} <$ 23 were slightly less likely to be disturbed, at 13 per cent (1/8), somewhat more likely to be relaxed, at 50 per cent (4/8), and moderately likely to be in an intermediate state, at 38 per cent (3/8). Unlike the low mass structures, there was no significant difference in the time since the last major merger. 

Like low-mass structures, high-mass structures with $N_{90, ICL} <$ 23 were also more likely to share the top progenitor of the ICL with the BCG. 25 per cent (2/8) shared the most significant progenitor, compared with 14 per cent (1/7) of those with $N_{90, ICL} >$ 69. The difference is less pronounced than that found for the lower-mass structures, possibly because major mergers play a larger role for the BCG in higher-mass structures, while the ICL is more likely to be formed from the accretion of satellite galaxies. For the high mass structures, with $N_{90, ICL} <$ 23 the fraction that the most significant progenitor made up of the ICL was $24 \pm 6$ per cent of the ICL, while high-mass structures with $N_{90, ICL} >$ 69 had a most significant progenitor fraction of $14 \pm 5$. The gap was larger for the fraction of the ICL composed of the two most significant progenitors, ranging from $42 \pm 10$ per cent to $23 \pm 7$ per cent. Like the low mass structures the number of progenitors affected the pathways by which the ICL forms in the high mass progenitors, however the effects were less pronounced. For high-mass structures with $N_{90, ICL} < $23, the percentage of the ICL that comes from progenitor galaxies that have completed their mergers was more significant, making up $35 \pm 16$ per cent, while those with $N_{90, ICL} > $69 had a completed mergers fraction of $23 \pm 9$ per cent. Correspondingly, high-mass structures with $N_{90, ICL} < $23 had a lower contribution of ICL originating in the stripping of still surviving galaxies, albeit within uncertainties of $55 \pm 18$ to $68 \pm 9$ for those with $N_{90, ICL} > $69. The in-situ component was not affected.

Broadly, we found that the results for high-mass structures were similar to those for low-mass structures, except for the dynamical states. $N_{90, ICL}$ generally made less of a difference for the high-mass structures, possibly due to more limited sampling or to these effects being less notable in high-mass clusters.

\citet{Brown2024} investigated ICL assembly in 11 clusters with masses between 1~$\times$~10\textsuperscript{14}~\(\textup{M}_\odot\) and 7~$\times$~10\textsuperscript{14}~\(\textup{M}_\odot\) in the Horizon-AGN simulation. They found that a few dozen progenitors typically form about 90 per cent of the ICL. Additionally, they find that a major accretion event (which disturbs a cluster) will form a large amount of ICL, although completed major mergers decrease the ICL fraction in clusters because they cause more growth in the BCG. Approximately one-third of the ICL stars were found to have previously belonged to the BCG, implying that violent BCG mergers could contribute to ICL growth.
Over longer timescales, the ICL is dominated by stars stripped from intermediate-mass contributors \citep[e.g.][]{Contreras_Santos_2024, mayes2025}. 

Our selection of structures with a few progenitors disproportionately favours those that have experienced a recent major merger. In contrast, the selection of structures with many progenitors favours those that have evolved over longer timescales and are dominated by the ongoing stripping of intermediate-mass galaxies. Massive satellites experience stronger dynamical friction and, therefore, deposit most of their material in the central galaxy \citep{Cooper2013}, which is why major mergers play a larger role in the growth of the BCG than the ICL.

Fig.~\ref{fig:BCGICLsigprogs}, plots the number of significant progenitors of the BCG against the number of significant progenitors of the ICL. We find that BCG significant progenitors scale with ICL progenitors, such that BCGs with a larger number of progenitors also have a larger number of progenitors in the ICL. The relation has a slope of 1.84, indicating that the BCG has, on average, fewer significant progenitors than the ICL. The number of significant progenitors in the BCG ranges from 1 to 50, with a mean of 10.2. \citet{Rodriguez_Gomez2016} investigated the stellar mass assembly of galaxies in the Illustris simulation and found that for BCGs, a large fraction of accreted mass is contributed by a small number of massive progenitors, with $N_{90}$ on the order of 10 for many structures, consistent with our results. This reflects the differing accretion histories of the BCG and the ICL: the ICL grows primarily from stellar and merger stripping of material from intermediate- and massive-sized galaxies, while the BCG grows in large part from major mergers, which are not expected to contribute significantly to the ICL \citep{demaio2015,mayes2025}. \citet{Kimmig_2025} compared the ICL+BCG fractions across the Magneticum, TNG100, Horizon-AGN and Hydrangea simulations and found that the ICL and the BCG were collectively dominated by a few massive progenitors that showed strong halo-to-halo scatter in $N_{90}$. While the number of significant progenitors to the BCG lags behind the ICL, Fig.~\ref{fig:BCGICLsigprogs} shows a clear correlation between the ICL and the BCG, indicating that the two co-evolve.

\begin{figure}

	\includegraphics[width=\columnwidth]{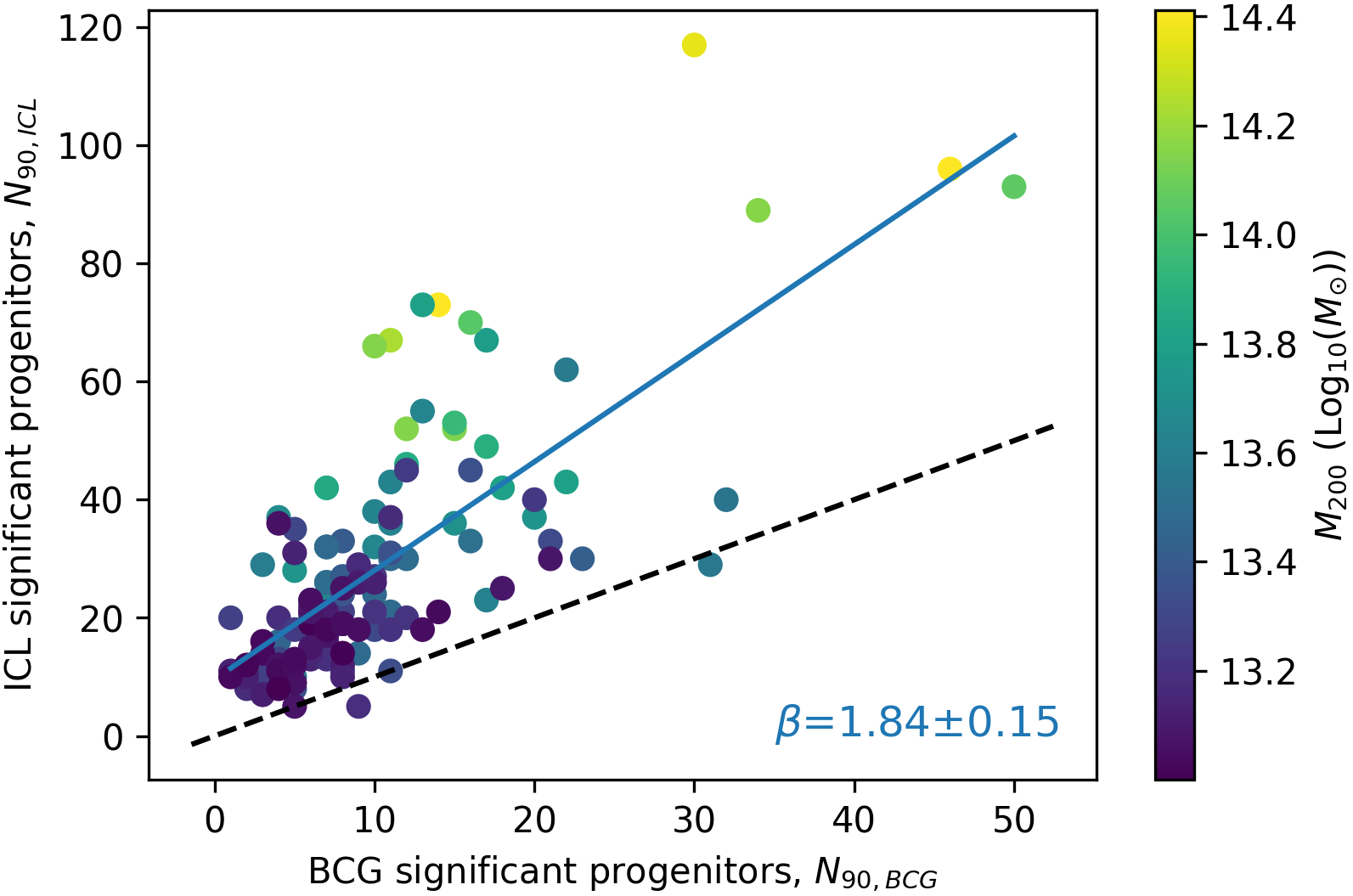}
    \caption{The number of significant progenitors that make up the ICL against the number of significant progenitors that make up the BCG, coloured by the mass of the host structure. The black dashed line is the one-to-one ratio.}
    \label{fig:BCGICLsigprogs}
\end{figure}

No dependence was found on structure state or ICL origin that could explain differences in the number of progenitors in the ICL and BCG. However, the plot shows that, for a fixed number of significant progenitors to the BCG, increasing the total structure mass increases the number of progenitors contributing to the ICL. This is likely because more massive clusters host a larger number of massive satellites \citep[e.g.][]{Yang2008, mayes2025}. Higher-mass clusters have a larger number of satellites that could be stripped relative to lower-mass structures. Additionally, more massive clusters have higher velocity dispersions, which lengthen dynamical friction timescales, allowing for a more extended period of tidal stripping for satellite galaxies. 

The mean stellar mass of the most significant progenitor was greater for the BCG than the ICL. The mean stellar mass for the most significant progenitor of the BCG was 1~$\times$~10\textsuperscript{11}~\(\textup{M}_\odot\), while the mean stellar mass for the most significant progenitor of the ICL was 5~$\times$~10\textsuperscript{10}~\(\textup{M}_\odot\). This reflects the fact that the ICL is generally built up by a large number of intermediate-mass galaxies undergoing stripping, whereas the BCG is built up by a small number of completed major mergers. 

In summary, major mergers are more likely to contribute to the growth of the BCG rather than the ICL. The most significant contributors to the BCG may be a small number of major mergers across low and high-mass structures, while structure mass plays a larger role for the number of progenitors in the ICL. Selecting structures with relatively fewer significant progenitors to the ICL may disproportionately favour those in which the ICL has been influenced by a recent major merger.

\section{Fraction of the ICL and BCG contributed by the top 5 progenitors}
\label{section:top5frac}

Accreted stellar halos and the ICL are generally assembled from a small number of massive progenitors \citep[e.g.][]{Cooper2010, Cooper20150}. In this section, we investigate the percentage of the ICL attributable to the top 5 most significant contributors, which together constitute the largest share of the ICL mass. 

Fig.~\ref{fig:top3conts} plots the fraction that the top 5 contributing progenitors, $N_{5}$, make up of the ICL. Note that this plot includes a random selection of 40 groups with masses less than 10\textsuperscript{13}~\(\textup{M}_\odot\), to test how this relation changes towards lower masses. The contribution of $N_{5}$ to the ICL above 10\textsuperscript{13}~\(\textup{M}_\odot\) ranges from 24 to 89 per cent with a mean fraction of $62 \pm 13$ per cent. The linear fit (over the >10\textsuperscript{13}~\(\textup{M}_\odot\) range) finds that the fraction decreases with structure mass on a slope of -0.22.  

We selected 15 structures with a mean percentage contribution to the ICL of $N_{5}$ $> 78$ per cent and 15 with a mean percentage $< 47$ per cent. The results were broadly similar to those found for $N_{90}$ in Section \ref{section:NICLprogs}, as would be expected since structures with a a large number of significant progenitors ($N_{90}$) will have a low contribution from the top 5 progenitors ($N_{5}$). Note, however, that for this selection we did not first restrict by structure mass, which will impact our results. The selection of structures with mean percentage $> 78$ per cent had a mean structure mass of 10\textsuperscript{13.2}~\(\textup{M}_\odot\), while the selection of structures with mean percentage $< 47$ per cent had a mean structure mass of 10\textsuperscript{14.0}~\(\textup{M}_\odot\). 

Halo state and concentration were unaffected by $N_{5}$. The mean lookback time since the last major merger for the structures with a high percentage of $N_{5}$ is $4.3 \pm 3.8$ Gyr, while structures with a low percentage of $N_{5}$ had $8.1 \pm 3.2$ Gyr, likely indicating that the structures with a high percentage have recently completed a major merger, which is still affecting the ICL. 

Structures with a high percentage of mass coming from the $N_{5}$ were more likely to share the most significant progenitor with the BCG. 53 per cent (8/15) shared the most significant progenitor, compared with only 13 per cent (2/15) of those with a low percentage from the top 5 contributors. In contrast to the $N_{90}$ results, the percentage from the top 5 progenitors had much less of an effect on the contribution from low mass (M\textsubscript{*}~<~1~$\times$~10\textsuperscript{10}~\(\textup{M}_\odot\)) progenitors. A large contribution from the top 5 contributors resulted in a low mass progenitor percentage of $25 \pm 8$ per cent, while a low contribution from $N_{5}$ gave $34 \pm 18$ per cent, which is within uncertainties. The difference between the results found from $N_{90}$ may be due to the fact that we did not place a structure mass range restriction on the selection of $N_{5}$. Similarly, there was minimal change in the mean progenitor mass. The percentage of the ICL coming from progenitor galaxies that have completed their mergers was $55 \pm 22$ per cent for the upper $N_{5}$ percentage and $29 \pm 13$ per cent for the lower $N_{5}$ percentage, which, unlike the results for $N_{90}$ is within uncertainties. Correspondingly, structures with a large $N_{5}$ had a percentage of ICL originating in the stripping of still surviving galaxies of $32 \pm 24$ per cent, while the structures with lower $N_{5}$ had $58 \pm 16$ per cent. 
The BCG generally had a higher percentage of material from $N_{5}$, ranging from 42 to 99 per cent with a mean percentage of $78 \pm 12$ per cent. This is likely due to major mergers having a larger influence on the growth of the BCG rather than the ICL.

\begin{figure}

	\includegraphics[width=\columnwidth]{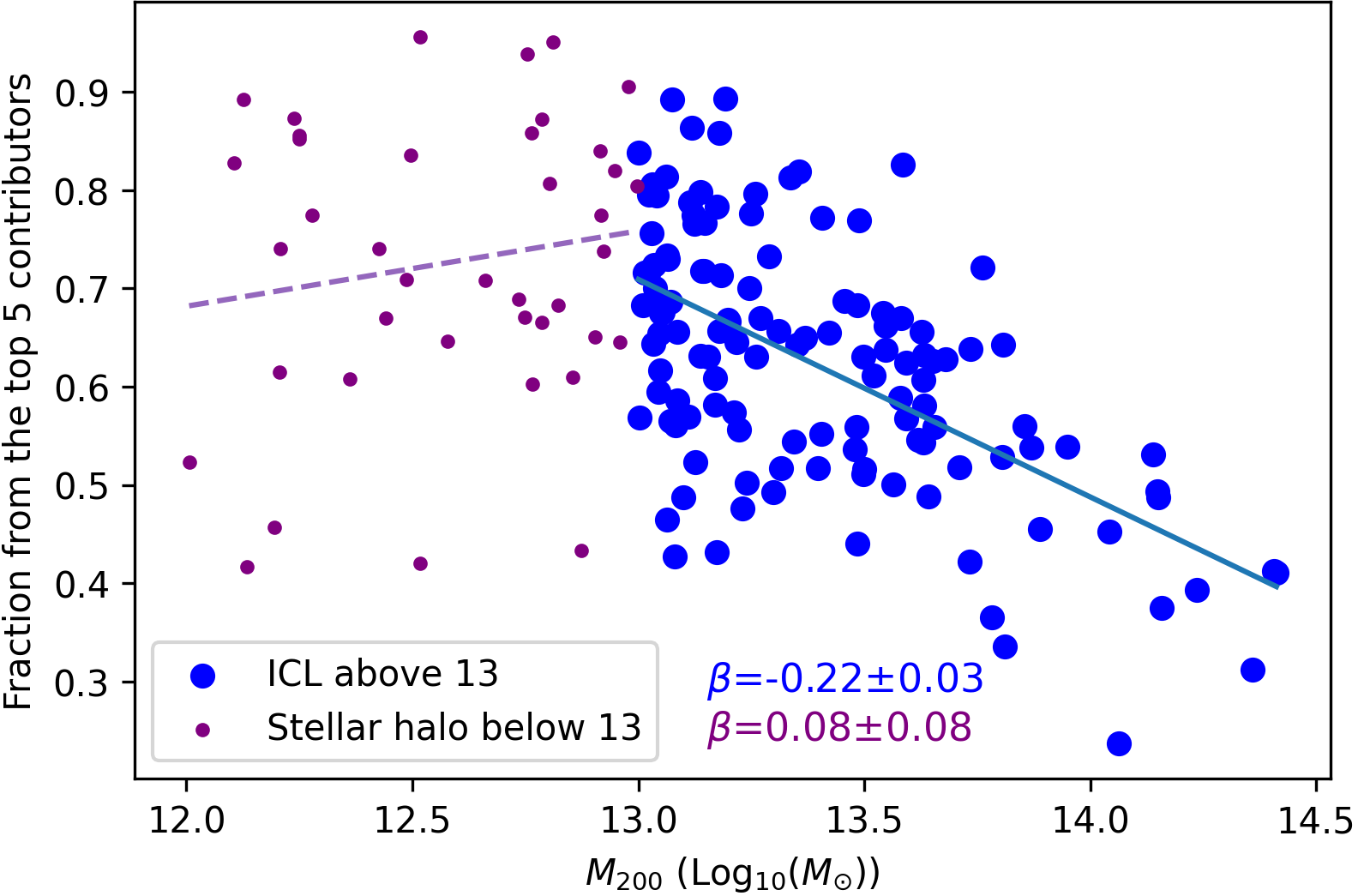}
    \caption{The fraction that the top 5 contributing progenitors make up of the ICL accreted material, including a small selection of sub 10\textsuperscript{13}~\(\textup{M}_\odot\) structures.}
    \label{fig:top3conts}
\end{figure}

\citet{Monachesi2019} studied the number of progenitors that contribute to the growth of Milky Way mass galaxy stellar halos using the Auriga simulations, and found that more massive accreted stellar halos generally had fewer significant progenitors. The stellar mass vs halo mass relation experiences a turn-over around peak halo mass of 10\textsuperscript{12}~\(\textup{M}_\odot\) \citep{Behroozi2019}, meaning the build-up of the ICL is different to the build-up of Milky Way mass galaxies.  To test this, we also include in Fig.~\ref{fig:top3conts} a random selection of 40 structures with mass less than 10\textsuperscript{13}~\(\textup{M}_\odot\). We select the stellar structure of these structures by the same method used to divide the ICL from the BCG in larger structures. Although there is substantial scatter, in part due to the small number of structures selected in this region, we find that the fraction of the stellar structure given by the top 5 contributors decreases with decreasing structure mass. 

The $L^*$ turnover occurs because more massive structures have a satellite mass function shifted toward higher masses \citep[e.g.][]{Yang2008}. The more massive progenitors (in total mass) contribute to the ICL a lower fraction of stellar mass than they would contribute if the turnover did not occur, which also results in a larger number of significant contributors. In Milky Way mass galaxies, more massive satellites bring more relative stellar mass, and therefore, the more massive galaxies are more significant. In the Milky Way mass range, the halos may host a few massive satellites that dominate the halos. A turnover occurs in the cluster mass range, where clusters begin to host numerous massive galaxies and satellites that are susceptible to stripping. Milky-Way mass galaxies generally have a much smaller ex-situ contribution than BCGs, of about 10 per cent and assemble early. In comparison, BCGs are dominated by ex-situ accretion and continue to accrete up to redshift 0 \citep{Rodriguez_Gomez2016}. \citet{Monachesi2019} and \citet{Fattahi_2020} both additionally found with Auriga that there is a strong radial dependence on the number of accreted galaxies contributing to Milky Way halos, with a few progenitors dominating the inner halo, and more contributing at higher radii, consistent with our finding that the ICL has more significant progenitors than the BCG, if we consider the ex-situ component of the BCG and the ICL as proxies for the inner and outer halo of Milky Way mass galaxies respectively.

\section{Contribution to the ICL of high mass progenitors}
In this section, we investigate the fraction of progenitors of stellar mass greater than 10\textsuperscript{9}~\(\textup{M}_\odot\) and 10\textsuperscript{11}~\(\textup{M}_\odot\) that make up the BCG and ICL.

Fig.~\ref{fig:sigprogmassICLBCG9}, plots the fraction that progenitors with stellar mass > 10\textsuperscript{9}~\(\textup{M}_\odot\) make up of the ICL and the BCG. The mass percentage that these progenitors make up of the BCG ranges from 93.2 to 99.2 per cent with a mean of $97.4 \pm 1.3$, indicating that the BCG is dominated by higher-mass galaxies over the chosen structure mass range, with little contribution from lower mass galaxies. There is little change with mass, with the linear fit having a slope of $0.009 \pm 0.003$, indicating a slight increase in fraction with structure mass. The ICL has a great deal more scatter, with the percentage ranging from 67.7 per cent to 97.0 per cent with a mean of $89.1 \pm 5.1$. The ICL percentage shows a clear dependence on structure mass, with a slope of $0.042 \pm 0.012$, indicating that higher mass structures have a higher percentage of accreted material from mass > 10\textsuperscript{9}~\(\textup{M}_\odot\) in the ICL. \citet{Brown2024} investigated the assembly of the ICL in structures with mass > 10\textsuperscript{14}~\(\textup{M}_\odot\) using the Horizon-AGN simulation and found that approximately 90 per cent was contributed from progenitors with mass > 10\textsuperscript{9}~\(\textup{M}_\odot\), consistent with our results.

\begin{figure}

	\includegraphics[width=\columnwidth]{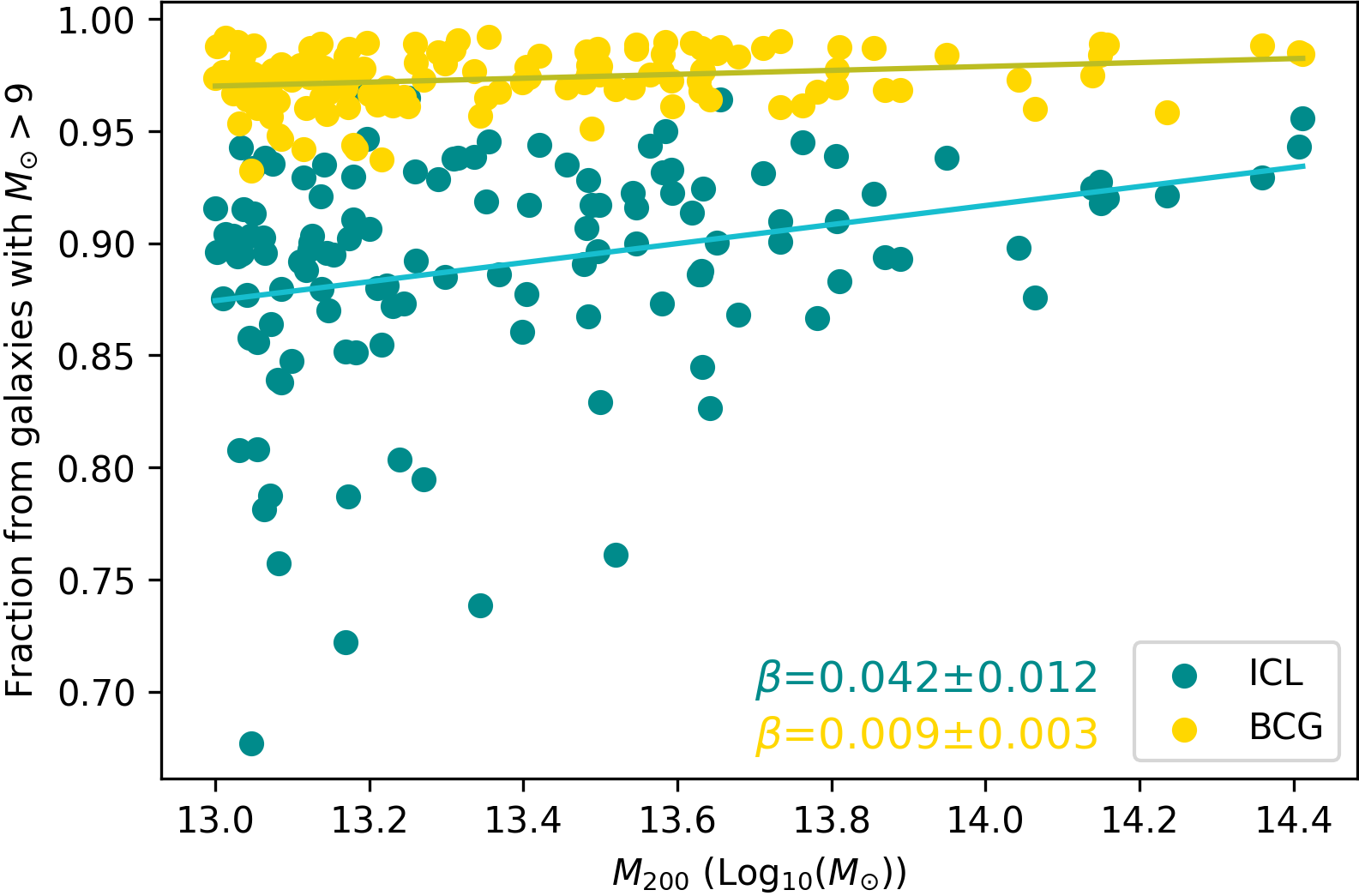}
    \caption{The fraction that progenitors with mass > 10\textsuperscript{9}~\(\textup{M}_\odot\) make up of the ICL and BCG}
    \label{fig:sigprogmassICLBCG9}
\end{figure}

The difference between the trends in the ICL and the BCG is likely due to the ICL being primarily composed of satellite galaxies that have experienced stripping. Because the ICL is dominated by the stripping of intermediate and high-mass galaxies the fraction contributed by massive galaxies increases more strongly with structure mass than the fraction in the BCG, as the cluster satellite mass function increases with mass.

We select 12 structures with a fraction that progenitors with mass > 10\textsuperscript{9}~\(\textup{M}_\odot\) make up of the ICL of greater than 94 per cent, and 12 structures with less than 82 per cent, to compare the effects that a high and low percentage of mass > 10\textsuperscript{9}~\(\textup{M}_\odot\) progenitors have on the ICL. Note that the low percentage is biased toward lower-mass structures, which may affect our results. The lower selection has a mean mass of 10\textsuperscript{13.2}~\(\textup{M}_\odot\), while the upper selection has a mean mass of 10\textsuperscript{13.8}~\(\textup{M}_\odot\).

Halos with a large percentage of mass > 10\textsuperscript{9}~\(\textup{M}_\odot\) progenitors were much more likely to be disturbed, with 50 per cent (6/12) disturbed, 17 per cent (2/12) relaxed, and 33 per cent (4/12) in an intermediate state, while those with a lower percentage of mass > 10\textsuperscript{9}~\(\textup{M}_\odot\) progenitors, were most likely to be in an intermediate state at 67 per cent (8/12), somewhat more likely to be relaxed at 33 per cent (4/12) and never disturbed at 0 per cent (0/12). This is likely related to total structure mass: structures with a large fraction of their mass from high-mass galaxies are in an active state of accretion.

Halos with a large percentage of mass > 10\textsuperscript{9}~\(\textup{M}_\odot\) progenitors were less concentrated at $4.86\pm 1.31$, while those with a low percentage (meaning a greater fraction of lower mass galaxies) had higher concentrations of $6.87\pm 1.79$, likely because less massive galaxies can be stripped in more concentrated structures. A higher percentage of massive progenitors indicated a more recent major merger at lookback time $4.7 \pm 3.6$ Gyr, while a lower percentage resulted in $8.2 \pm 2.8$ Gyr. More concentrated structures are more evolved \citep{Sereno_2013}. Merging is more efficient in these structures and, thus, more likely to contribute to the growth of the BCG than the ICL.

A large percentage of massive galaxies had little impact on whether the dominant formation pathway of the ICL was progenitor galaxies that had completed mergers or were still surviving and being stripped. However, structures with a large percentage of massive galaxies had smaller in-situ components, at $9.8 \pm 4.5$ to $20.7 \pm 4.7$. Note that while the completed mergers and surviving progenitors formation pathways are often separated, they are ultimately different stages of the same pathway, therefore the in-situ material can be used as a general assessment of accretion.

Fig.~\ref{fig:sigprogmassICLBCG11}, plots the fraction that progenitors with mass > 10\textsuperscript{11}~\(\textup{M}_\odot\) make up of the ICL and the BCG. Both the ICL and the BCG show a lot of scatter. The BCG had percentages ranging from 78.0 to 0 per cent with a mean percentage of $12.8\pm 21.7$. The ICL had percentages ranging from 83.4 to 0 per cent with a mean percentage of $11.4\pm 19.1$. The percentage for both increases with structure mass with the BCG increasing on the slope of 0.35, and the ICL on the slope of 0.29. The BCG slope is slightly steeper, indicating a stronger increase with mass, although the slopes are within uncertainties. The large scatter is likely because in this mass range, whether a cluster or group has accreted a galaxy with a stellar mass of > 10\textsuperscript{11}~\(\textup{M}_\odot\) is highly variable, with many structures likely only accreting a single galaxy in this mass range, and others accreting none, especially at lower masses. 

In all 67 structures where the ICL had no contribution from 10\textsuperscript{11}~\(\textup{M}_\odot\) progenitors, the BCG contribution was also 0. Out of the 72 structures where the BCG had no contribution from 10\textsuperscript{11}~\(\textup{M}_\odot\) progenitors 5 structures had a minimal contribution from progenitors with mass > 10\textsuperscript{11}~\(\textup{M}_\odot\) in the ICL, making up less than 2 per cent of the ICL fraction, possibly due to a massive galaxy just entering the structure or passing by. \citet{Brown2024} found that roughly half the ICL comes from galaxies with stellar masses within 0.5 dex of 10\textsuperscript{11}~\(\textup{M}_\odot\).
\citet{Contreras_Santos_2024} investigated the formation of the ICL in the three hundred simulations, and found that a large fraction of approximately 65-80 of the accreted ICL comes from progenitor galaxies with mass > 10\textsuperscript{11}~\(\textup{M}_\odot\), in structures with masses greater than 8~$\times$~10\textsuperscript{14}~\(\textup{M}_\odot\). Both studies investigated high-mass progenitor galaxies in high-mass structures. While we found a large variation in whether or not the ICL contained material from mass > 10\textsuperscript{11}~\(\textup{M}_\odot\) progenitor galaxies towards the group environment, towards higher mass clusters our results are consistent with these studies.

\begin{figure}

	\includegraphics[width=\columnwidth]{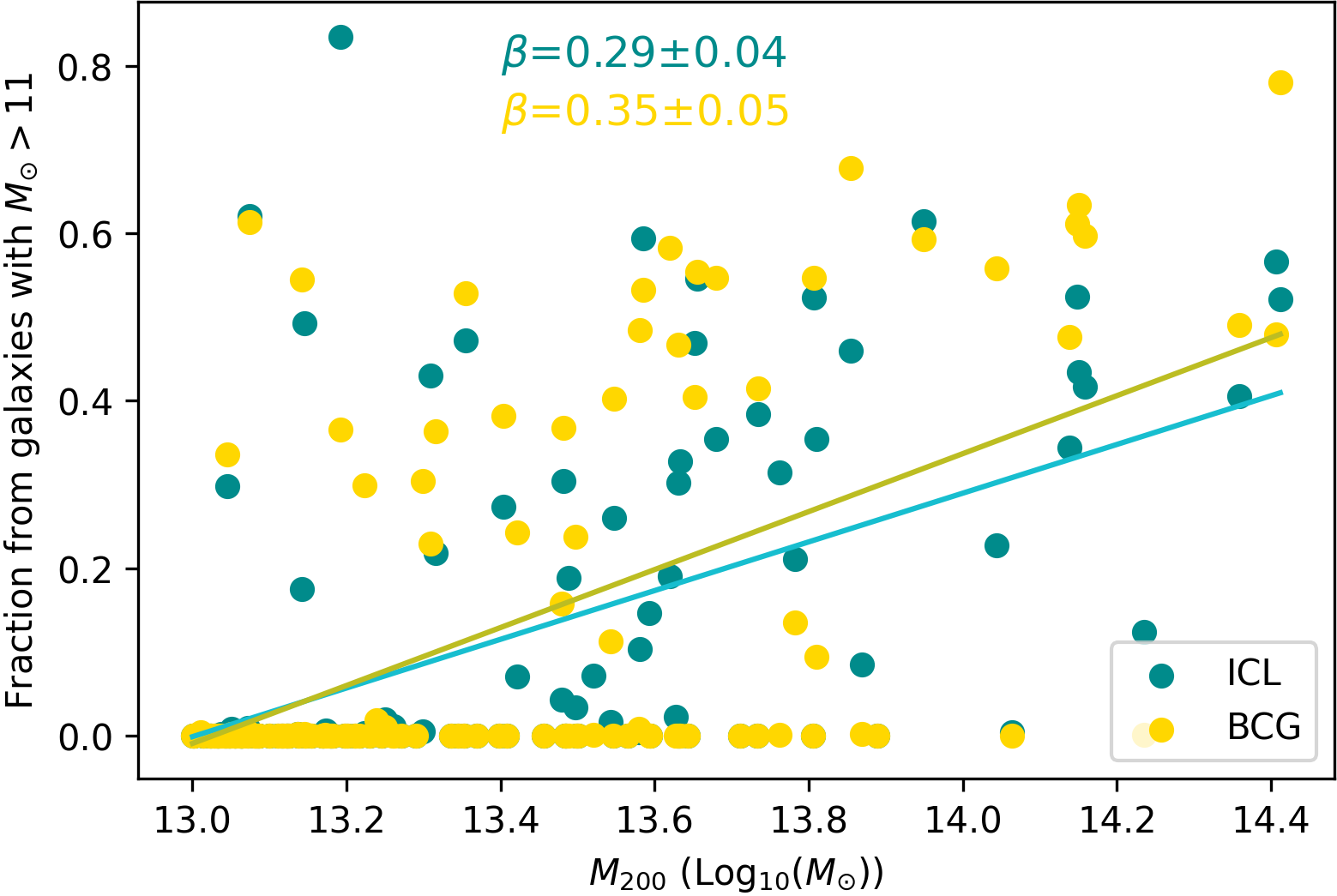}
    \caption{The fraction that progenitors with mass > 10\textsuperscript{11}~\(\textup{M}_\odot\) make up of the ICL and BCG}
    \label{fig:sigprogmassICLBCG11}
\end{figure}

\section{Contribution to the ICL from preprocessed progenitors}
\label{section:Preprocessing}
In this section, we investigate preprocessed ICL, that is, ICL formed from progenitor galaxies quenched before entering the cluster. Fig.~\ref{fig:preprocessed} plots the fraction of accreted material that is from quenched (preprocessed) galaxies. The fraction of the ICL from quenched progenitors increases with structure mass, as would be expected, as lower mass structures may still be assembling. \citet{Pallero2019} studied the quenching of galaxies in clusters in the EAGLE simulation and found that more massive clusters accrete greater numbers of quenched galaxies. The majority of galaxies become quenched within structures with total masses between 10\textsuperscript{13.5}~\(\textup{M}_\odot\) and 10\textsuperscript{14.5}~\(\textup{M}_\odot\). Thus, a substantial amount of satellites in our lower mass structures will still be experiencing quenching. 

The ICL has a mean fraction of material associated with processed galaxies, $f_{\rm pp}$, of $36.7 \pm 19.3$ per cent, ranging from 1.6 to 78.1 per cent, while the BCG $f_{\rm pp}$ has $31.0 \pm 22.4$ per cent, ranging from 0.1 to 89.3 per cent. The ICL $f_{\rm pp}$ has a slope of 0.24, while the BCG $f_{\rm pp}$ has a slope of 0.13, indicating that the ICL progenitors are more likely to be quenched than the BCG progenitors as structure mass increases. 

\begin{figure}

	\includegraphics[width=\columnwidth]{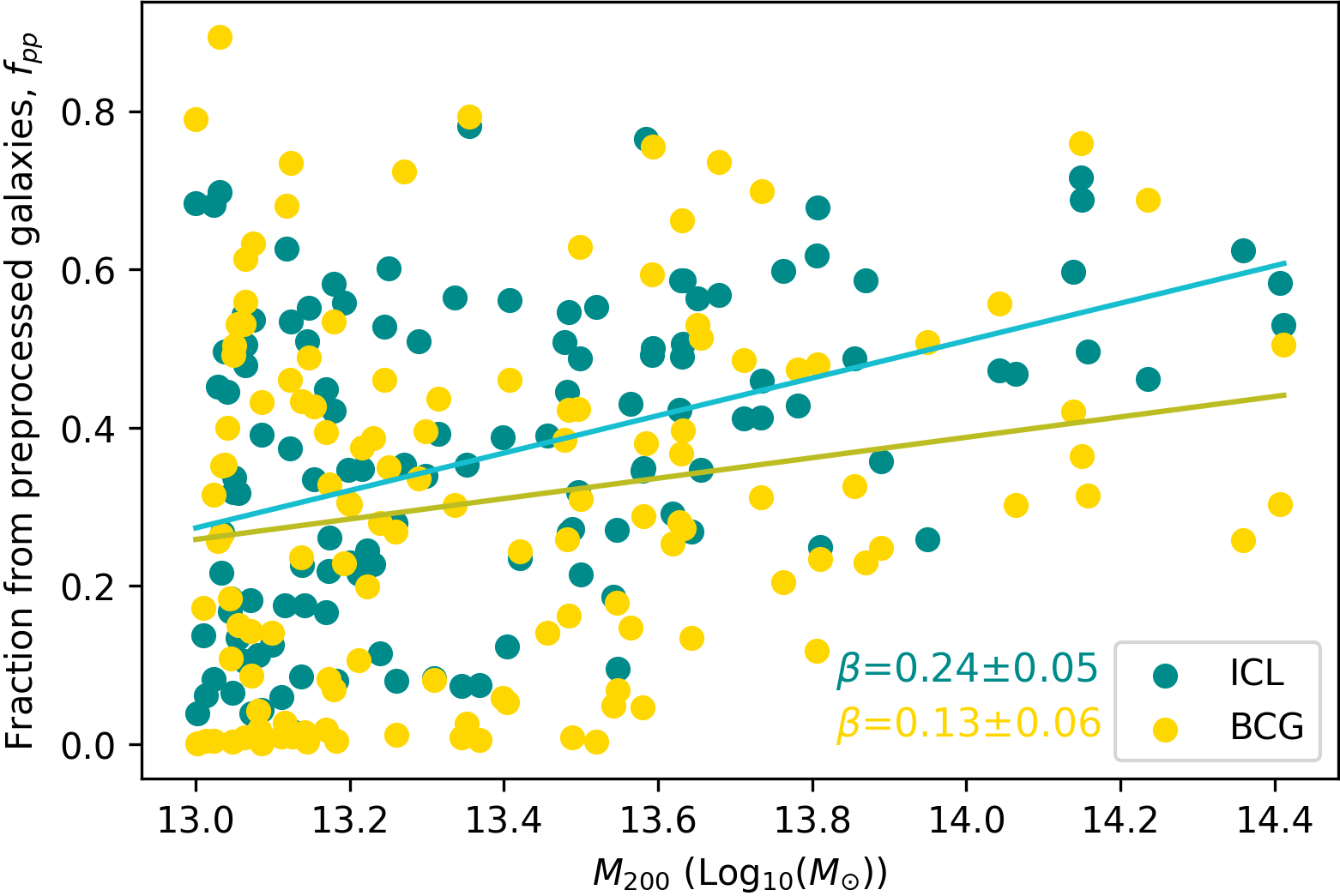}
    \caption{The fraction of accreted material from preprocessed galaxies}
    \label{fig:preprocessed}
\end{figure}

To investigate the differences between systems where progenitor galaxies are quenched we selected 23 structures with a $f_{\rm pp}$ greater than 56 per cent and 23 with a $f_{\rm pp}$ less than 36 per cent. The structures with a higher $f_{\rm pp}$ were slightly more likely to be disturbed than relaxed, with 30 per cent (7/23) of the structures being disturbed, 26 per cent (6/23) being relaxed, and most structures being in an intermediate state at 43 per cent (10/23). The lower selection were less likely to be disturbed than relaxed at 17 per cent (4/23) disturbed and 26 per cent (6/23) being relaxed, and more likely to be in an intermediate state at 57 per cent (13/23). The upper fraction was significantly less concentrated at $5.4 \pm 1.2$ to $6.9 \pm 1.0$. Time since the last major merger was not found to be affected.

\citet{Chun2023} applied the galaxy replacement technique to 6 clusters with M~>~1~$\times$~10\textsuperscript{14}~\(\textup{M}_\odot\) using multi-resolution cosmological N-body simulations, while the follow-up work \citet{Chun2024} did the same for 84 structures in the mass range of 4~$\times$~10\textsuperscript{13}~\(\textup{M}_\odot\) - 6~$\times$~10\textsuperscript{14}~\(\textup{M}_\odot\). They found that half the progenitors of the ICL in unrelaxed clusters had been pre-processed before entering the host structure, slightly higher than our results, and contributing primarily to the outer regions of the clusters. While we found that structures with a higher quenched fraction were more likely to be disturbed, they were equally likely to be relaxed, distinct from \citet{Chun2023}'s results.

Unsurprisingly, since $f_{\rm pp}$ is correlated with structure mass it is also correlated with the fraction that low mass progenitors made up of the ICL, with the upper selection having a low mass percentage (M\textsubscript{*}~<~1~$\times$~10\textsuperscript{10}~\(\textup{M}_\odot\)) of $24 \pm 7$ and the lower selection having a low mass fraction of $50 \pm 21$. The mean progenitor mass for the upper selection was $10.8 \pm 0.2$~\(\textup{M}_\odot\), while the lower selection had $10.2 \pm 0.3$~\(\textup{M}_\odot\). $f_{\rm pp}$ also impacted the percentage of ICL mass coming from completed mergers and surviving satellites, albeit within uncertainties. The upper percentage showed a completed mergers ratio of $35 \pm 20$ per cent, and surviving $54 \pm 23$ per cent, while the lower percentage had a completed mergers ratio of $49 \pm 16$ per cent, and surviving $32 \pm 15$ per cent. 

ICL observations have also found evidence of pre-processing in galaxy clusters. \citet{Ellien_2025} observed the intracluster light of Abell 2390, and estimated that around 30 per cent of the ICL was diffuse light brought in by smaller groups.
\citet{Jimnez_Teja_2025} using deep view imaging of the Coma cluster, found features in the ICL that indicate they were likely still bound to small groups at discovery.

Different simulation studies may define ICL preprocessing differently, which affects our comparisons with them. \citet{contini2014} used semi-analytic models to study ICL in 10\textsuperscript{13}~\(\textup{M}_\odot\)- 10\textsuperscript{15}~\(\textup{M}_\odot\) mass structures and found that between 5-25 per cent of the ICL was accreted while the structures were assembling, slightly lower than our percentages. \citet{Pallero2019} find that 10-20 per cent of galaxies are quenched before infall into low mass ($\approx$10\textsuperscript{14}~\(\textup{M}_\odot\)) clusters, but 50-70 per cent have undergone some form of preprocessing. 
\citet{Brown2024} studied the assembly of the ICL in 11 clusters with mass > 10\textsuperscript{14}~\(\textup{M}_\odot\) in the Horizon-AGN Simulation. They defined preprocessed ICL as star particles that entered the cluster unassociated with a progenitor galaxy. They found a wide range in preprocessed percentages, from 8-44 per cent.

\section{Progenitors mean mass gradient}
\label{section:mmassslope}

In this section, we investigate the change in the progenitor mass gradient of the BCG+ICL and the ICL. For each structure, stellar particles from the BCG+ICL are placed in 30 radial bins from the centre of the structure to $R_{200}$, with the restriction that each bin must contain at least 5 particles. The mean mass of the progenitor galaxies for all ex-situ stellar particles is calculated within each bin and plotted. The data is then fitted with a linear function. The linear functions for all structures are plotted in Fig.~\ref{fig:massslope}. The fitting process is then repeated separately for the ICL alone. Appendix Fig.~\ref{fig:massslopeex} plots examples of linear function fitting for 6 randomly selected individual structures. 

\begin{figure}

	\includegraphics[width=\columnwidth]{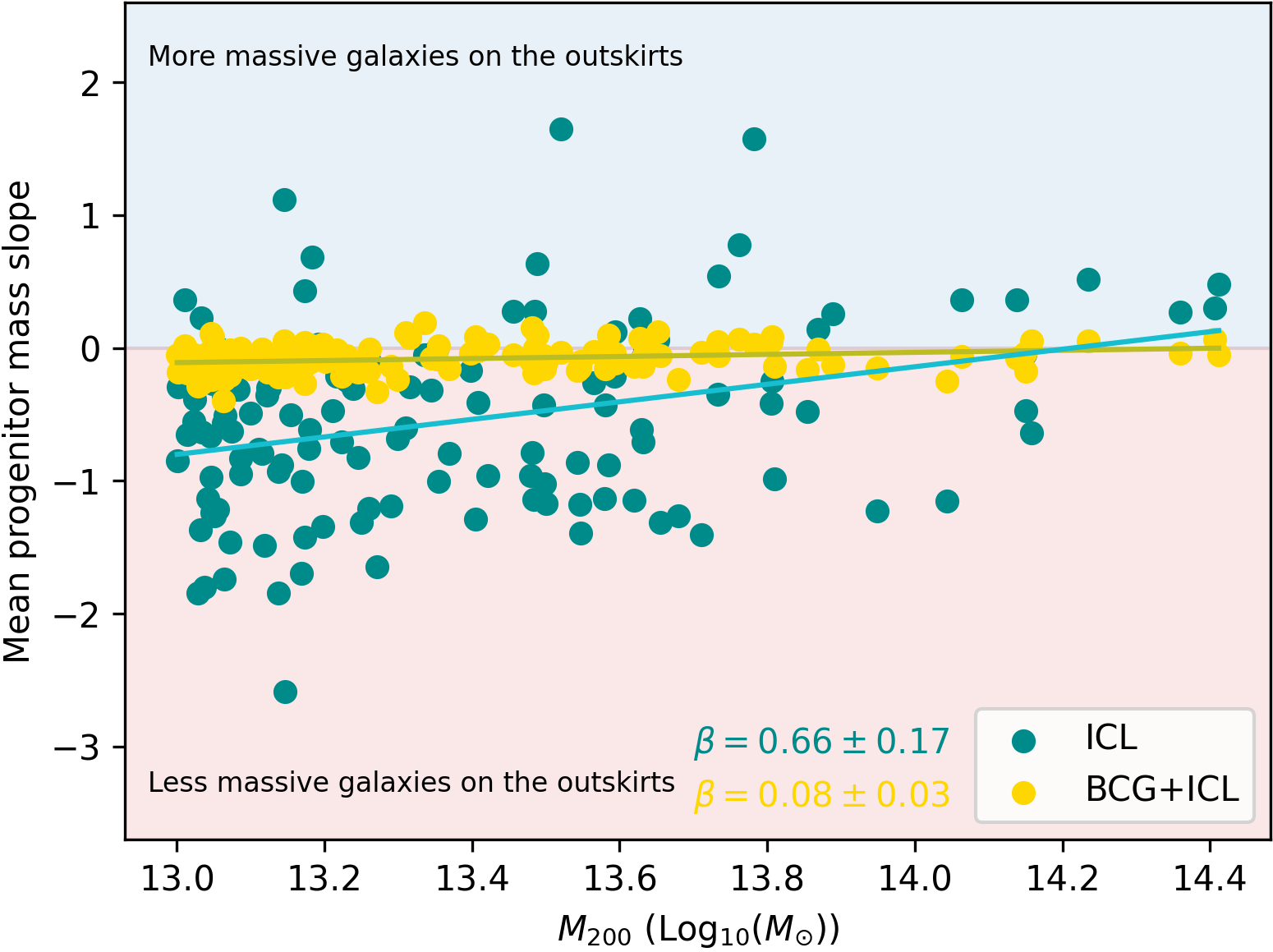}
    \caption{Halo linear functions of the mean mass gradient of the progenitor galaxies that formed the BCG+ICL.}
    \label{fig:massslope}
\end{figure}

The BCG+ICL mean mass gradient slope was largely flat, with a slope of $0.08 \pm 0.03$, indicating a small positive gradient (more massive galaxies on the outskirts) for more massive structures. There was little scatter in the BCG+ICL relation, indicating little variation between structures. The overall slope is dominated by the BCG, so we do not consider plotting the BCG gradient separately.

The ICL mean mass gradient slope was positive with a slope of $0.66 \pm 0.17$, indicating that towards higher masses, structures have less negative slopes. This indicates that the masses of progenitor galaxies on the outskirts of the ICL increases relative to the masses of progenitor galaxies found in the inner ICL, possibly due to ongoing mergers. However, the majority of structures (101 of 127) have negative slopes, indicating that material from more massive progenitors is generally found towards the inner ICL. There is a great deal of scatter in the slopes of the ICL mean mass gradient, ranging from -2.6 to 1.7, with a mean of $-0.54 \pm 0.70$. The formation of ICL in structures is known to be highly individualistic, depending on the structure history, which is shown by these plots. 

Other simulation studies show similar results. \citet{Ahvazi2024} used the TNG50 simulations and found that the inner regions of the ICL are more likely to have higher-mass progenitors. \citet{contini2014} modelled the formation of the ICL and BCG semi-analytically and found that the BCG core forms by a few major mergers, while the ICL is dominated by massive satellites at intermediate radii and smaller satellites at larger radii, which would give a negative radial mass gradient. They also found that pre-processing and group infall can deliver material from massive progenitors to the outer regions. In observations \citet{Montes2018} imaged cluster cores using the Hubble Frontier Fields and found that the inner ICL is consistent with having been built by massive progenitors, based on the high metallicities found in this region, whereas the outskirts show greater variation.

Unsurprisingly, the ICL mass gradient significantly affected structure properties. We select 19 structures with slope greater than 0.22, and 19 structures with slope less than -1.24 for comparison. We find that the mass slope has a large impact on whether a structure is disturbed or relaxed with 42 per cent (8/19) of the positive slope structures being disturbed and 16 per cent (3/19) being relaxed. In contrast, only 11 per cent (2/19) of the negative slope structures were disturbed, while 47 per cent (9/19) were relaxed. Concentration was also significantly impacted, with the positive slope structures being less concentrated at $4.86 \pm 1.58$ and the negative slope structures having $7.09 \pm 1.17$. Lookback time since the last major merger for the positive slope structures was $7.0 \pm 4.1$ Gyr, and $4.4 \pm 2.6$ Gyr for the negative slope structures. The negative slope structures are therefore highly evolved, while the positive slope structures are typically in an active state of disruption. The lookback time since the last major merger would seem to contradict this; however, note that this statistic is only for completed major mergers. It does not factor in the possibility that structures with positive slopes are likely actively undergoing or are about to undergo a major merger, which is probable for a structure that has not experienced a recent completed major merger \citep[e.g.][]{Amorisco2017}.

The mass gradient impacted whether the structures shared the most significant progenitor between the ICL and the BCG. For positive slope structures (more massive galaxies on the outskirts), 26 per cent (5/19) shared the most significant progenitor, while 58 per cent (11/19) shared the most significant progenitor with the BCG for the negative slope. The mass gradient also significantly affected the formation pathways of the surviving and completed merger components. The positive gradient structures had completed mergers component $29.8 \pm 16.4$ per cent and surviving component $57.2 \pm 19.3$ per cent, while the negative gradient structures had completed mergers component $57.3 \pm 18.1$ per cent and surviving component $27.0 \pm 19.1$ per cent. This is likely because the negative slope structures are more evolved, while the positive slope structures are more likely to be experiencing active accretion in the ICL region of the structure. The in-situ percentages were not affected, nor was the fraction of low mass progenitors. \citet{Kimmig_2025} compared a variety of simulations (MAGNETICUM, TNG, Hydrangea, Horizon-AGN) and found that the inner ICL is dominated by massive progenitors from early major mergers, while the outer ICL shows more variation and is sensitive to recent group accretion. They find that the radial progenitor mix depends on the cluster's dynamical state. Relaxed clusters were smoother with a more centrally concentrated progenitor distribution, while unrelaxed clusters showed greater variation, especially at high radii.

\section{Conclusions}
\label{section:conc}
In this paper, we use the hydrodynamic IllustrisTNG-100 simulation to investigate the progenitor galaxies of the ICL. We select 127 clusters and groups with masses >10\textsuperscript{13}~\(\textup{M}_\odot\), and separate the ICL from the BCG with an r-band surface brightness cut at the Holmberg radius of 26.5 mag/arcsec\textsuperscript{-2}. In this section, we summarise our results.

\subsection{Number of significant progenitors}

The number of ICL progenitors increases with structure mass and consistently exceeds that of the BCG, although both populations co-evolve. Furthermore, structures with a large number of ICL progenitors tend to be in intermediate dynamical states (lacking recent major mergers) and assemble their ICL primarily from completed mergers of low-mass galaxies, which typically do not overlap with the main progenitors of the BCG.

\subsection{Fraction of the top 5 progenitors}

The contribution of the top 5 progenitors to the ICL exhibits a distinct turnover at a structure mass of approximately $10^{13} M_{\odot}$. For Milky Way-sized and smaller systems, this fraction increases with structure mass, but for more massive systems, the trend reverses, and the top 5 contribution steadily declines. Additionally, systems heavily dominated by their top 5 contributors typically experienced a recent major merger and share their primary progenitor with the BCG, contrasting with the BCG's overall tendency to be shaped by fewer, more massive progenitors

\subsection{High mass progenitors}

Both the ICL and the BCG are predominantly assembled from massive progenitors ($>10^9 M_{\odot}$), with this contribution to the ICL increasing slightly alongside structure mass. Systems heavily enriched by these high-mass progenitors are typically in an active state of disruption, characterised by disturbed dynamics, lower concentrations, recent major mergers, and smaller in-situ stellar components.

\subsection{Preprocessing}

Preprocessed (quenched) progenitors contribute significantly to the accreted material of both the ICL (roughly 37 per cent) and the BCG (roughly 31 per cent), with the ICL's quenched fraction increasing in more massive structures. Structures characterised by a high fraction of preprocessed ICL progenitors tend to be dynamically disturbed and less concentrated; additionally, their ICL is primarily stripped from surviving, higher-mass galaxies rather than being built from completed mergers.

\subsection{Mean mass gradient}
Although the majority of structures exhibit a negative ICL mean mass gradient, the gradient tends to become positive in higher-mass clusters, indicating that material from massive progenitors is deposited on the outskirts rather than the centre. These positive-slope systems are characteristically dynamically disturbed and less concentrated, as their ICL is primarily fed by ongoing stripping from surviving galaxies rather than recently completed mergers. Consequently, their ICL rarely shares its most significant progenitor with the BCG.

\bigskip  
\noindent
The goal of this paper was to determine the properties of the progenitor galaxies of the ICL and to compare them with those of the BCG. The ICL assembles from a large number of intermediate-mass progenitors, in contrast to the BCG, which has a few massive progenitors. Progenitor properties can vary strongly for a given structure mass, with the assembly of the ICL being highly individual and dependent on a structure's formation history. Future work should focus more specifically on how the ICL may build up and evolve over time, as this work focused exclusively on redshift 0 structures.

\section{Acknowledgments}

\begin{acknowledgements}
RJM and FAG acknowledge ANID Fondo 2021 GEMINI ASTRO21-0061, and Concurso Fondo ALMA-ANID 2024-31240052. The authors gratefully acknowledge support from the ANID BASAL project FB210003. FAG and AM acknowledge support from the ANID FONDECYT Regular grants 1251493 and 1251882, respectively, as well as funding from the HORIZON-MSCA-2021-SE-01 Research and Innovation Programme under the Marie Sklodowska-Curie grant agreement number 101086388. 
\end{acknowledgements}

\bibliographystyle{aa}
\bibliography{biblio} 

\appendix
\onecolumn

\section{Example plots of BCG+ICL mass linear functions}

\begin{figure}[ht]
\centering
\begin{multicols}{3}
    \includegraphics[width=\linewidth]{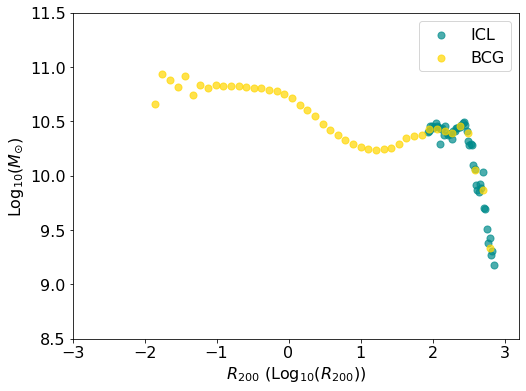}\par 
    \includegraphics[width=\linewidth]{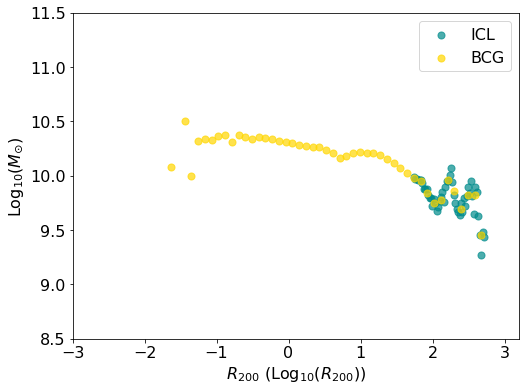}\par 
    \includegraphics[width=\linewidth]{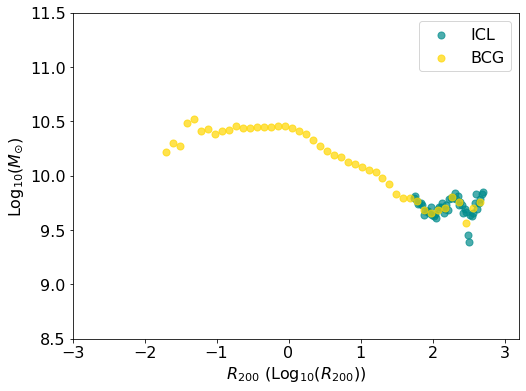}\par 
    \end{multicols}
\begin{multicols}{3}
    \includegraphics[width=\linewidth]{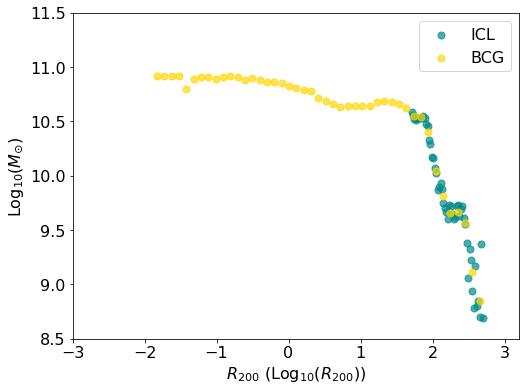}\par
    \includegraphics[width=\linewidth]{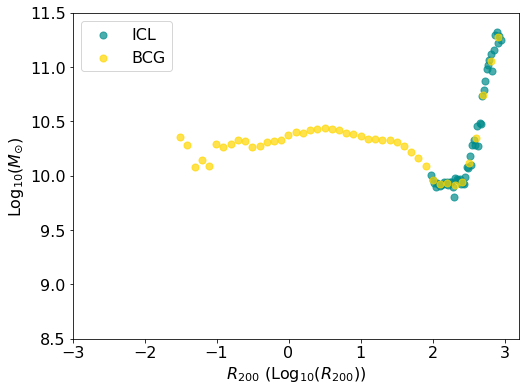}\par
    \includegraphics[width=\linewidth]{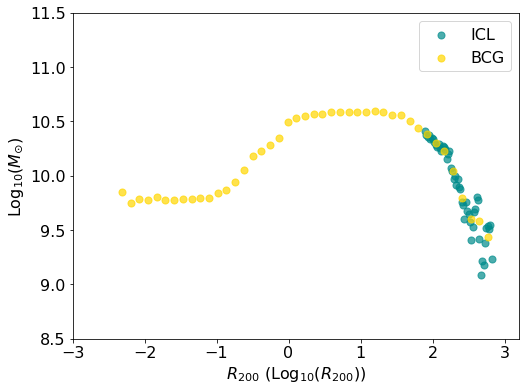}\par 
\end{multicols}
\caption{Example plots of BCG+ICL mass linear functions}
\label{fig:massslopeex}
\end{figure}

\end{document}